\documentclass{article}
\usepackage{spconf,amsmath,graphicx,hyperref}
\usepackage{booktabs}
\usepackage{multirow}
\usepackage{caption}
\usepackage{amssymb}
\usepackage{subcaption}
\usepackage{xcolor}

\title{Towards Real-Time Generative Speech Restoration with Flow-Matching}
\twoauthors
 {Tsun-An Hsieh}
	{University of Illinois Urbana-Champaign\\
	Urbana, IL, USA}
 {Sebastian Braun}
	{%Microsoft Corporation\\
	Microsoft Research\\
	Redmond, WA, USA}
\begin{document}
\ninept
\maketitle
\begin{abstract}
Generative models have shown robust performance on speech enhancement and restoration tasks, but most prior approaches operate offline with high latency, making them unsuitable for streaming applications. In this work, we investigate the feasibility of a low-latency, real-time generative speech restoration system based on flow-matching (FM). Our method tackles diverse real-world tasks, including denoising, dereverberation, and generative restoration. The proposed causal architecture without time-downsampling achieves introduces an total latency of only 20 ms, suitable for real-time communication. In addition, we explore a broad set of architectural variations and sampling strategies to ensure effective training and efficient inference. Notably, our flow-matching model maintains high enhancement quality with only 5 number of function evaluations (NFEs) during sampling, achieving similar performance as when using ~20 NFEs under the same conditions. Experimental results indicate that causal FM-based models favor few-step reverse sampling, and smaller backbones degrade with longer reverse trajectories. We further show a side-by-side comparison of FM to typical adversarial-loss-based training for the same model architecture. \footnote{Audio examples are available at \href{https://www.notion.so/TOWARDS-REAL-TIME-GENERATIVE-SPEECH-RESTORATION-WITH-FLOW-MATCHING-2721d2d22ec180a28582c1f67d9b1cee?source=copy_link}{our demo page}.}
%GAN training tends to improve speech fidelity while leaving more residual noise. Improvements are dataset- and input-quality dependent, with larger gains on heavily degraded signals.
\end{abstract}
\section{Introduction}
\label{sec:intro}
% Speech enhancement (SE) \cite{loizou2007speech} aims to estimate clean speech from noisy and/or reverberant mixtures. Classical methods, such as spectral subtraction \cite{lim1962enhancement} and MMSE \cite{ephraim2003speech, griffin1984signal}, typically rely on strong statistical assumptions, which often lead to suboptimal performance and may leave unnatural residual noise at low SNR \cite{o2024speech}. Modern deep neural network (DNN) systems learn masks (e.g., IRM \cite{narayanan2013ideal}, PSM \cite{erdogan2015phase}, and cIRM \cite{williamson2015complex}), complex spectral components \cite{tan2018convolutional}, or waveforms \cite{luo2019conv} directly, dropping many of these assumptions and consistently outperforming handcrafted pipelines, especially when generalizing to unseen noise types and acoustic environments.
Speech enhancement (SE) estimates clean speech from noisy or reverberant mixtures \cite{loizou2007speech}. Classical methods rely on strong statistical assumptions and often leave artifacts at low signal-to-noise ratio (SNR) \cite{o2024speech}. In contrast, modern DNNs learn masks \cite{narayanan2013ideal,erdogan2015phase,williamson2015complex}, complex spectra \cite{tan2018convolutional}, or waveforms \cite{luo2019conv}, relax these assumptions, and consistently outperform traditional pipelines.

Despite the success of predictive methods, SE is traditionally defined to address a limited set of problems--noise suppression and dereverberation. In contrast, speech restoration (SR) broadens beyond additive and convolutive degradations to include generative restoration tasks such as packet loss concealment, bandwidth extension, codec artifact removal, and declipping. These are often missing-data problems that require generating plausible content (e.g., filling lost packets or absent high-frequency content). 
To this end, early studies explored using the generative adversarial network (GAN) and diffusion models for SR. 
\cite{hu2024general} proposed a two-stage cascade network comprising a restoration module for SR and an enhancement module for conventional SE, jointly optimized with multiple discriminators.
% \cite{fu2021metricgan+} directly takes the noisy spectrum, uses an enhancement network to generate the enhanced speech, and improves the quality score estimated by the discriminator in an adversarial manner. 
% Diffusion models have also shown stronger generalization to unseen conditions than predictive models and have effectively extended SE to SR. 
\cite{lemercier2023analysing} argues that for non-additive SR tasks, learning a speech prior with generative models is preferable to predictive (regression/masking) approaches, and \cite{richter2023speech} empirically demonstrates superior generalization on unseen test data. Previous studies put emphasis on denoising diffusion probabilistic models \cite{ho2020denoising} and score-matching \cite{song2020score}. For example, \cite{lu2022conditional} integrates diffusion into SE by conditioning the reverse process on the noisy speech, while \cite{welker22speech} leverages denoising score matching to map noisy to clean speech. Nevertheless, without reverse-process corrections \cite{lay2024single}, diffusion-based SE models typically require 30-60 NFEs at inference, constraining sampling efficiency. All but one submission to the SIG challenge \cite{ristea2024icassp} with under real-time constraints are GAN based, indicating under-exploration of new methods for this task. 

Recent studies show that flow-matching (FM)–based SE can significantly improve sampling efficiency, reaching few-step or even one-step reverse process. \cite{lee2025flowse} reduces the NFE down to 5 and outperforms a diffusion counterpart requiring 60 NFEs. \cite{jung2024flowavse} adopts a two-stage cascade of predictive and generative modules. \cite{liu2024generative} and \cite{ku2025generative} fine-tune a pre-trained generative model for tasks such as speech separation, text-to-speech, and speech restoration. \cite{korostik2025modifying} exhaustively explores training/sampling strategies across tasks. 
These studies establish the feasibility of FM for SR; however, they are built on non-causal architectures, limiting suitability for real-time and streaming use. \cite{richter2024causal} presents a causal variant of \cite{welker22speech} that relies only on past information, yet its temporal downsampling substantially increases latency, rendering it unsuitable for real-time applications.

In this work, to our knowledge as the first in literature, we exploit FM architectures with a low algorithmic latency suitable for real-time communication. Unlike \cite{richter2024causal}, which despite being causal, has a large algorithmic latency of over 0.6~s due to heavy temporal downsampling in the U-net, we downsample only in frequency to reduce total latency to 20~ms. 
In addition, we exploit the effect of a significantly less complex architecture than typically used in diffusion or FM, and use this to directly compare GAN and FM training paradigm using the largely identical architecture. 
%consists of convolutional layers and gated linear units (GLUs) \cite{dauphin2017language}. 
% For evaluation, we train these models on a large-scale dataset with publicly available subsets and test them on the blind test sets of Deep Noise Suppression (DNS) Challenge \cite{dubey2022icassp} and Speech Signal Improvement (SIG) Challenge \cite{ristea2024icassp}. 
%
We evaluate our models by training on large-scale data with high amount of augmentation to cover all possible degradations and help generalization. We evaluate on the established real-world SIG and DNS challenge test sets to assess realistic performance expectations. 
%
% The experimental results provide a comprehensive view. First, regardless architectural differences and latency, FM models peak at varying but low number of NFEs (between 2 and 10), while too long reverse trajectories can degrade speech quality and word error rate (WER). Second, compared with GAN training, FM-based SR requires larger models and more compute to parameterize time-dependent vector fields. Third, the methods show complementary biases: GAN-based models emphasize speech clarity at the expense of more high-frequency artifacts, while FM emphasizes noise suppression, potentially at the expense of residual speech distortion.

% \section{Generative Speech Restoration}
% \label{sec:gsr}

\section{Methodology}
\subsection{Speech Restoration}
\label{ssec:prob_def}
% Our models operate in the compressed complex STFT domain to mitigate the heavy-tailed distribution of STFT spectrograms \cite{richter2024causal}. Given a complex spectrogram $c \in \mathbb{C}^{F\times L}$, where $F$ and $L$ denote the numbers of frequency bins and frames, respectively, the amplitude compression is defined as $\tilde{c} \;=\; \beta\,|c|^{\alpha} e^{i\angle c}$, where $\alpha$ is a compression exponent; $\beta$ is a scaling factor; $|\cdot|$ and $\angle(\cdot)$ denote the amplitude and phase operators, respectively.

% SR aims to recover clean speech from its degraded form. Let $x_1\in\mathbb{C}^{F\times L}$ denote the clean speech and $y\in\mathbb{C}^{F\times L}$ the degraded observation. SR assumes a general degradation operator that may compose multiple effects in arbitrary orders; we consider additive noise, reverberation, bandwidth truncation, packet loss, clipping, and codec distortions.

Let $x_1 \in \mathbb{C}^{F\times L}$ denote the clean STFT spectrogram and $y \in \mathbb{C}^{F\times L}$ a degraded observation. We assume $y$ is generated by an unknown composition of degradations in arbitrary order,
\begin{equation}
y = \mathcal{D}(x_1),
\qquad
\mathcal{D} = \mathcal{D}_K \circ \cdots \circ \mathcal{D}_1,
\end{equation}
where each $\mathcal{D}_k$ may represent additive noise, reverberation, bandwidth truncation, packet loss, clipping, or codec distortions. The goal of SR is to recover a plausible clean spectrogram $\hat{x}_1$ that is consistent with $x_1$.

\subsection{Flow-Matching for Real-time Speech Restoration}
\label{ssec:fm_for_sr}

Conditional FM \cite{lipman2022flow, tong2024improving} specifies a time-indexed conditional probability path $p_t(x | x_1)$, $t\in[0,1]$, that interpolates between a simple base at $t=0$ and the data at $t=1$. The path is realized by a probability-flow ODE whose flow map $\phi_t$ transports the base distribution $p_0$ to $p_t$:
\begin{equation}
\frac{d}{dt}\phi_t(x| x_1) = v_t\big(\phi_t(x| x_1)\big),
\qquad \phi_0(x| x_1)=x_0,
\end{equation}
where $v_t$ is the (time-dependent) target velocity field associated with the chosen path $p_t$. Training regresses a neural field $v_\theta$ toward this target at samples $x_t\sim p_t$:
\begin{equation}
\mathcal{L}_{\mathrm{CFM}}
=\mathbb{E}_{x_1}\int_0^1 \mathbb{E}_{x_t\sim p_t}
\big|
v_\theta (x_t, t) - v_t(x_t| x_1)
\big|^2 dt,
\end{equation}
and sampling integrates $dx_t = v_\theta(x_t,t)\,dt$ from $x_0\sim p_0$ to obtain a draw at $t=1$.

Given paired data $\{(x_0,x_1)\}$ (noisy $x_0$, clean $x_1$), we model $p(x_1| x_0)$ using a \textit{Gaussian conditional path} with time-specific mean and standard deviation
\begin{equation}
\mu_t = tx_1,\qquad
\sigma_t = (1-t)\sigma_{\mathrm{max}} + t\sigma_{\mathrm{min}},
\end{equation}
so that $p_t(x| x_1)=\mathcal{N}(x;\mu_t,\sigma_t^2 I)$ transitions from a broad base around the origin at $t=0$ to a sharp distribution near $x_1$ as $t\to 1$ (deterministic at $t=1$ if $\sigma_{\mathrm{min}}=0$). This choice yields the closed-form target vector field
\begin{equation}
u_t(x| x_1)=
\frac{\sigma_{\mathrm{max}}x_t - \sigma_{\mathrm{min}}(x_t-x_1)}
{(1-t)\sigma_{\mathrm{max}}+t\sigma_{\mathrm{min}}}.
\end{equation}
We condition the estimator on the observation only through its inputs, $v_\theta(x_t,x_0,t)$, and train with the OT-CFM objective
\begin{equation}
    \mathcal{L}_{\mathrm{OT\text{-}CFM}} = 
    \mathbb{E}_{(x_1, x_0),t, x_t}
    \Vert
    v_\theta(x_t, x_0, t) - u_t(x_t|x_1)
    \Vert_2^2,
\end{equation}
where the clean/noisy pair $(x_1, x_0)$ is sampled from the data distribution $p_{\mathrm{data}}$; $t\sim\mathcal{U}(0,1)$, and $x_t\sim p_t=\mathcal{N}(x_t;\mu_t, \sigma_t)$.
At inference, we integrate the ODE from $t=0$ to $1$, producing $x_1$ conditioned on $x_0$.

\subsection{Model Architectures}
\label{ssec:model}
\subsubsection{Causal NCSN++}
Our causal model architecture is built on the causal encoder–decoder of \cite{richter2024causal} with no temporal downsampling. The backbone is a U-net with skip connections in which all temporal operations use causal convolutional layers and employ cumulative group normalization, without look-ahead. Multi-scale context is obtained only along frequency via stride-2 downsampling at each stage and symmetric upsampling in the decoder. With five scales, this yields an overall 32$\times$ frequency pyramid while keeping the temporal stride at 1 end-to-end. Consequently, the temporal receptive field is expanded by depth rather than by temporal striding. The FM time step $t$ is embedded with Gaussian Fourier projection followed by a multi-layer perceptron, and the resulting time embedding is injected into each layer.

\subsubsection{ConvGLU-UNet}
As a simpler, lower complexity alternative, we use ConvGLU-UNet, which is a simple 1D UNet made up of convolutional GLUs. A GLU has two parallel convolutions, one as a linear path and one as a gate, where we use Tanh activation, which better suits complex audio data being a symmetric signal. One conv block is made up as depthwise separable convolution, i.\,e.\, a grouped depthwise followed by a 1x1 pointwise conv.  The encoder uses depthwise separable conv layers with kernel size 2 while the decoder use only plain 1x1 GLUs without temporal operations. The bottleneck is a ConvGLU with kernel size 7. The conv channels of encoder are (4096,2048, 1024,512,256,128) and mirrored in the decoder. We add linear skip connections with 1x1 mappings from each encoder feature to the corresponding decoder feature.  

\section{Experimental Settings}
\label{sec:exp}

\subsection{Datasets}
\label{ssec:datasets}

\subsubsection{Preprocessing}
Our models operate in the compressed complex STFT domain to mitigate the heavy-tailed distribution of STFT spectrograms \cite{richter2024causal}. Given a complex spectrogram $c \in \mathbb{C}^{F\times L}$, where $F$ and $L$ denote the numbers of frequency bins and frames, respectively, the amplitude compression is defined as $\tilde{c} = \beta\,|c|^{\alpha} e^{i\angle c}$, where $\alpha$ is a compression exponent; $\beta$ is a scaling factor; $|\cdot|$ and $\angle(\cdot)$ denote the amplitude and phase operators, respectively.

\begin{table}[tb]
  \centering
  \footnotesize
  \setlength{\tabcolsep}{3pt}        % tighten column padding
  \renewcommand{\arraystretch}{0.95} % tighten row spacing
  \caption{Model complexity. RF denotes the length of receptive field.}
  \label{tab:model_comp}
  \begin{tabular*}{\columnwidth}{@{\extracolsep{\fill}} lccc}
    \toprule
    \textbf{Model} & \textbf{Params (M)} & \textbf{MACs/s (G)} & \textbf{RF (s)} \\
    \midrule
    Non-causal         & 53.0 & 65.69  & 3.82 \\
    Causal             & 53.0 & 142.78 & 0.53 \\
    ConvGLU-UNet-base  & 6.02 & 0.36   & 0.75 \\
    ConvGLU-UNet-large & 57.6 & 3.5    & 0.75 \\
    \bottomrule
  \end{tabular*}
\end{table}

\subsubsection{Training \& Test Sets}
The training data is generated on the fly and follows the training data used in \cite{braun2022effect}.
Speech and noise data is from the DNS challenge \cite{dubey2022icassp}, where the speech is filtered for high MOS and SNR. We add random degradations to simulate various acoustic effects, using bandwidth limitation with various upper and lower cutoffs and filter types, non-linear distortion with various and randomly parameterized distortion types, codec artifacts from GSM and MP3, random masking of time-frequency patches, and level variations. The target signal is kept at a target level of -25 dBFS, so that the model learns to output a constant level, as also used in \cite{richter2024causal}.

% \subsubsection{Test Sets}
% DNS Challenge 2022 \cite{dubey2022icassp} blind testing emphasizes large-scale, real-recorded denoising in communication scenarios. It contains 859 fixed 10-second clips at 48 kHz (we resample to 16 kHz), crowd-sourced across desktop and mobile with nearly 70\% smartphone recordings, and each clip pairs a unique speaker with a unique background noise to preclude speaker/noise repetition and stress generalization in noise suppression. In SIG Challenge 2024 \cite{ristea2024icassp}, the blind set consists of 500 clips, shared across real-time and non-real-time tracks to enforce cross-track comparability. Each clip comes from a distinct device, environment, and speaker, recordings span PCs and mobile devices and multiple languages, and the set is explicitly stratified over impairment categories (i.e., noise, reverberation, and other distortions), reflecting a broader signal improvement remit beyond pure denoising.
We adopt the blind test sets from DNS Challenge 2022 \cite{dubey2022icassp} and SIG Challenge 2024 \cite{ristea2024icassp}. For better clarity, we term them as DNS and SIG2024 in the remainder of the paper to distinguish with the MOS socres. DNS comprises 859 real 10-second clips at 48 kHz (we resample to 16 kHz), crowd-sourced across desktop and mobile with about 70\% smartphone recordings, each pairing a unique speaker and background noise to promote denoising generalization. The SIG2024 data contains 500 clips; each clip comes from a distinct device, environment, and speaker across PCs/mobile and multiple languages, and the set is stratified over impairment types (noise, reverberation, other distortions) to reflect a broader signal-improvement remit. In short, DNS focuses on noise suppression, whereas SIG emphasizes broader signal improvement and generative restoration.

\subsection{Evaluation Metrics}
\label{ssec:metrics}

We use non-intrusive MOS estimators to evaluate the audio quality, given that the generative restoration task allows a multitude of good solutions. To complementarily monitor the content preservation, we use automatic speech recognition (ASR) \cite{radford2023robust} and compute word error rate (WER) \cite{morris2004and}.

DNSMOS \cite{reddy2021dnsmos} is trained directly on ITU-T P.835 subjective ratings and outputs three perceptual dimensions per clip: speech quality (SIGMOS), background noise quality (BAKMOS), and overall quality (OVRLMOS), that mirror the P.835 protocol. 
%The model consumes log-power spectrogramscomputed at 16 kHz with 20 ms frames and 10 ms hop, and a CNN regresses the three MOS scores. 
DNSMOS shows high correlation with human judgments, and is trained to rank noise suppressors. DNSMOS generalization to other degradations as in SIG2024 is however less strong, so its results on SIG2024 have to be taken with care.
DistillMOS \cite{stahl2025distillation} also predicts non-intrusively overall MOS, but has been trained on a wide variety of data and degradations, including all degradations considered in SIG. The degradations include VoIP, generative artifacts, clipping, distortion, bandwidth limitation, on-device processing such as speech enhancement, audio coding etc. DistillMOS distills a large self-supervised, XLS-R–based teacher into compact students using a large compilation of MOS-labeled datasets plus pseudo-labels on unlabeled, realistically degraded speech. Notably, the authors observe that DNSMOS performs well on the DNS Challenge 3 \cite{reddy2021interspeech} set but generalizes less well to several other datasets, motivating broader-trained MOS predictors such as DistillMOS.

%Using DNSMOS together with the broadly generalizing DistillMOS, gives both fine-grained diagnostics, such as speech distortion, residual noise, and a cross-dataset-robust perceptual summary, yielding a more reliable, reference-free evaluation of our SR models on real-world test sets.

% The MOS prediction reflects perceived naturalness and quality, but it does not capture whether the restored speech preserves the original linguistic content. WER provides a complementary measure since it evaluates intelligibility by comparing the words in the restored speech with the ground-truth transcript. This distinction is important because a model may generate speech that sounds clean and natural, yet if the spoken content is altered or distorted, the WER remains high and reveals the limitation of relying on MOS estimators alone.

\begin{figure*}[tb]
    \centering
    \includegraphics[width=0.93\textwidth]{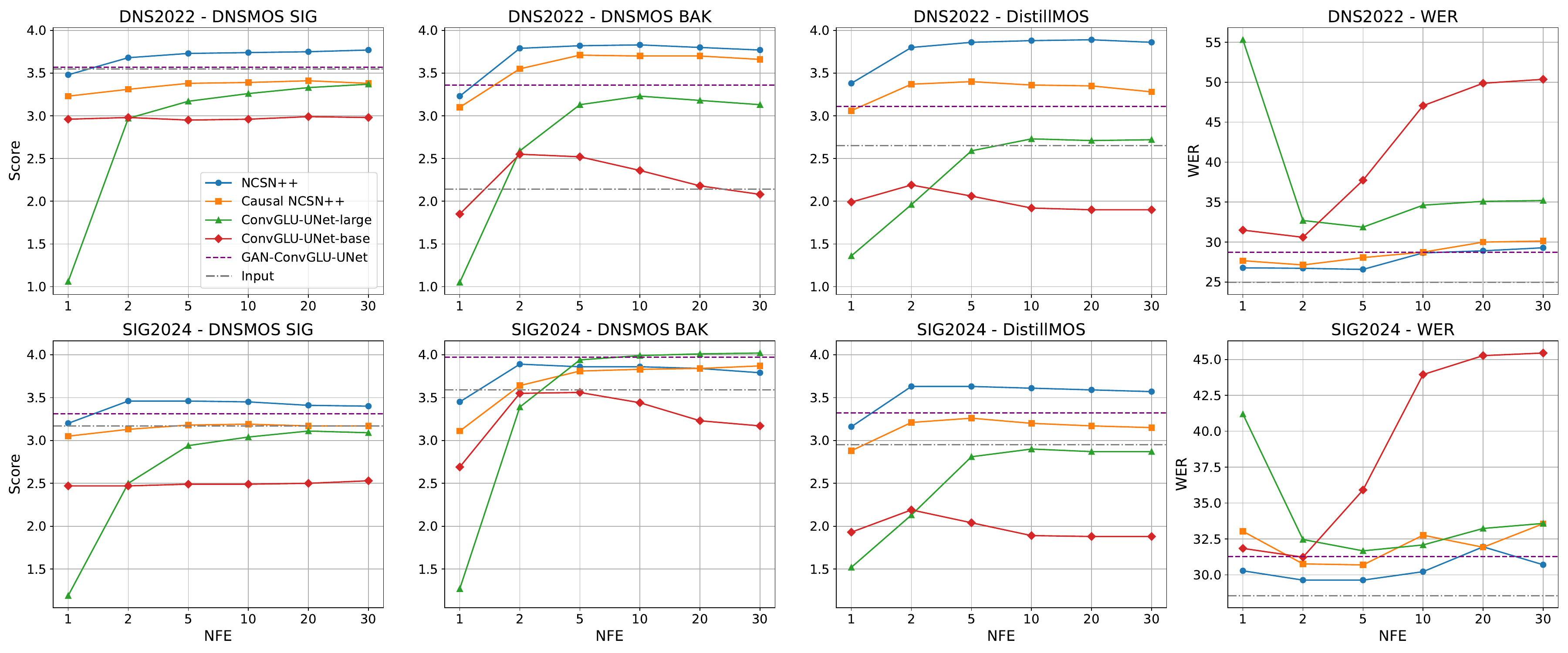}
    \caption{Average scores by NFEs across DNS and SIG datasets.}
    \label{fig:performance}
    \vspace{-10pt}
\end{figure*}

\subsection{Model Complexity}
\label{ssec:complexity}
Table \ref{tab:model_comp} contrasts model capacity, compute, and receptive field (RF). The non-causal and causal NCSN++ have the same number of parameters but very different profiles. The non-causal model leverages look-ahead and temporal down/up-sampling, yielding a large RF of 3.82 s and a compute of 65.69 G MACs/s, whereas our causal variant removes future context and keeps full time resolution, which shrinks RF to 0.53 s but more than doubles throughput cost to 142.78 G MACs/s. This trade-off keeps algorithmic latency low at the expense of compute and long-range context.
The ConvGLU-UNets illustrate efficiency scaling. The base model is computationally two orders of magnitudes cheaper than NSCN++, while the large large variant is still far cheaper than NSCN++ at 3.5 G MACs/s with the same RF. No attempts of such significant down-scaling model complexity for speech processing via FM have been reported in literature so far. 

% \begin{figure*}[t]
%   \centering
%   % Row 1: DNS 2022
%   \begin{subfigure}[t]{0.32\textwidth}
%     \centering\includegraphics[width=\linewidth]{figs/dnsmos_sig_nfe_DNS2022.png}
%     \caption{DNS'22: DNSMOS SIG}
%   \end{subfigure}\hfill
%   \begin{subfigure}[t]{0.32\textwidth}
%     \centering\includegraphics[width=\linewidth]{figs/dnsmos_bak_nfe_DNS2022.png}
%     \caption{DNS'22: DNSMOS BAK}
%   \end{subfigure}\hfill
%   \begin{subfigure}[t]{0.32\textwidth}
%     \centering\includegraphics[width=\linewidth]{figs/distillmos_nfe_DNS2022.png}
%     \caption{DNS'22: DistillMOS}
%   \end{subfigure}

%   \vspace{0.6em}

%   % Row 2: SIG 2024
%   \begin{subfigure}[t]{0.32\textwidth}
%     \centering\includegraphics[width=\linewidth]{figs/dnsmos_sig_nfe_SIG2024.png}
%     \caption{SIG'24: DNSMOS SIG}
%   \end{subfigure}\hfill
%   \begin{subfigure}[t]{0.32\textwidth}
%     \centering\includegraphics[width=\linewidth]{figs/dnsmos_bak_nfe_SIG2024.png}
%     \caption{SIG'24: DNSMOS BAK}
%   \end{subfigure}\hfill
%   \begin{subfigure}[t]{0.32\textwidth}
%     \centering\includegraphics[width=\linewidth]{figs/distillmos_nfe_SIG2024.png}
%     \caption{SIG'24: DistillMOS}
%   \end{subfigure}

%   \caption{Two-by-three grid: top row = DNS'22 scores; bottom row = SIG'24 scores.}
%   \label{fig:two_by_three_doublecol}
% \end{figure*}[t]

\begin{figure}[t]
    \centering
    \includegraphics[width=\linewidth,clip,trim=20 0 0 0]{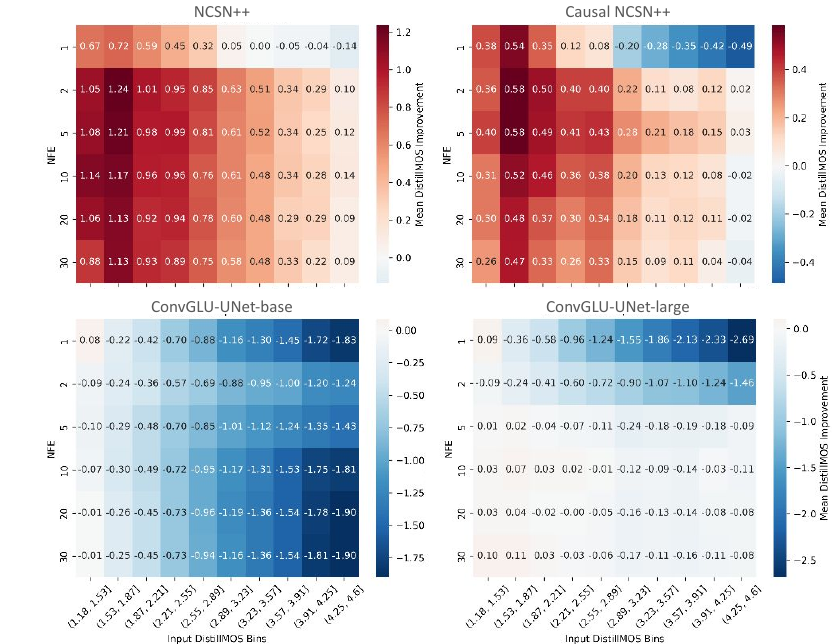}
    \caption{DistillMOS improvement comparison by NFE and input quality on SIG testset.}
    \vspace{-10pt}
    \label{fig:breakdown}
\end{figure}
\section{Results}
\label{sec:nfe_break}

\subsection{Performance Comparison of FM-based Models}

Figure~\ref{fig:performance} shows across both datasets, the non-causal NCSN++ establishes the upper bound baseline with the highest SIGMOS and DistillMOS, with most of the gain realized by NFE = 2 and only marginal changes thereafter. The causal NCSN++ follows the same trend but sits consistently lower, approximately 0.3 on SIGMOS and 0.5 on DistillMOS. The ConvGLU-UNet family behaves differently with respect to NFE. The base model does not improve with additional steps; its DistillMOS decreases after 5 NFEs. The large model improves sharply from NFE = 1 to NFE = 5 and then plateaus. On both DNS and SIG it achieves SIGMOS comparable to the causal NCSN++, while offering a similar model size and roughly 40 times lower compute. 

BAKMOS trends diverge from SIGMOS and DistillMOS. Under the noisier DNS condition, the non-causal model is slightly better than the causal model; under the less noisy condition, as NFE increases, the causal model narrows the gap with the non-causal model. For the ConvGLU-UNets, the large model requires more NFEs to reach strong noise suppression (on the SIG dataset it reaches around 4.0 and even edges past the non-causal NCSN++), whereas the base model is best at NFE between 2 and 5; for the base model, BAKMOS peaks around NFE = 2 and then degrades as NFE increases.
WER largely mirrors the BAKMOS behavior with a few differences. For the ConvGLU-UNet-base, WER increases as NFE increases, which can indicate increasing hallucination. The large model approaches the NCSN++ models after 2 NFEs (whereas its BAKMOS requires 5 NFEs), and the WER gap between the causal and non-causal NCSN++ models remains small without significant differences.

Overall, the results suggest that low capacity models combined with long ODE trajectories tends to over-regularize or accumulate error. FM on very small backbones is not powerful enough to model complex speech, and running at too many steps can introduce artifacts that further degrade the signal. In contrast, operating in a few-step regime (2 to 5) is consistently preferable for ASR compatibility, with high-capacity models benefiting the most.

\subsection{GAN vs. FM-based Training}
As a baseline, we train ConvGLU-UNet also with typical STFT reconstruction and adversarial losses. We term this model GAN-ConvGLU-UNet. The discriminator setup largely follows the multi-resolution discriminators in \cite{defossez2022highfi}.
The GAN-ConvGLU-UNet baselines (horizontal red lines in Fig.~\ref{fig:performance}) sit between causal and non-causal NCSN++ on SIGMOS and below both NCSN++ models on DistillMOS.
Higher SIGMOS than FM on the same ConvGLU backbones but lower BAKMOS, i.e., better speech retention with more residual noise. On DNS the GAN’s overall perceptual score is below both NCSN++ models but above the ConvGLU-FM variants; on SIGMOS it surpasses the causal NCSN++ in DistillMOS yet still trails the non-causal NCSN++. Finally, the input baselines (grey lines) are higher on SIGMOS than on DNS, so absolute headroom is smaller. The only model able to improve SIGMOS for DNS is the non-causal NSCN++. Given the recent fast success of diffusion and FM, it is surprising that under the same architecture and runtime constraints, FM models do not seem to be preferable to established GAN models.

% but the model ranking and NFE trends remain consistent: use non-causal NCSN++ at NFE=5--10 for best quality, causal NCSN++ at similar NFEs when streaming is required, and GAN-ConvGLU when you prefer bright and crisp speech over maximal noise suppression; FM on ConvGLU-large can push BAKMOS very high with enough steps but does not close the gap on SIGMOS/overall quality.

\subsection{DistillMOS Breakdown}
Fig.~\ref{fig:breakdown} presents a breakdown of DistillMOS improvements as a function of the NFEs and input quality for the SIG dataset, with warmer colors indicating larger gains. Overall tendencies are consistent across architectures, i.e. improvements increase as the input degrades, and for high-quality inputs the scores first rise with NFE, reach a model-specific maximum, and then decrease monotonically with further steps, except for ConvGLU-UNet-large.

The optimal NFE is stable across input quality and differs by causality. The non-causal model peaks at 10 NFEs, whereas the causal variant peaks at 5 NFEs on both datasets. At low input quality, the non-causal model delivers roughly twice the improvement of the causal model, while at high input quality their gap narrows substantially. This pattern suggests that FM benefits more from severe contamination and should be run near its dataset-agnostic optimum rather than pushed to longer trajectories.

Across ConvGLU-UNet variants, the score distributions mirror those of causal NCSN++, with performance saturating at relatively low NFE. For low-quality inputs, however, ConvGLU-UNet-base requires substantially higher NFEs to reach near-undegraded performance. By contrast, the ConvGLU-UNet-large continues to gain as the NFE increases, indicating that the FM trades off larger model capacity with longer reverse trajectory.

In summary, \ref{fig:breakdown} supports three practical guidelines: (i) select NFE by architecture (about 10 for non-causal NCSN++, 5 for causal NCSN++); (ii) expect larger gains on heavily degraded inputs and smaller gains on cleaner inputs; and (iii) avoid over-processing past the peak NFE, which yields unwanted artifacts and increases the chances of hallucination.

\begin{figure}[t]
    \centering
    \includegraphics[width=\linewidth,clip,trim=0 10 0 10]{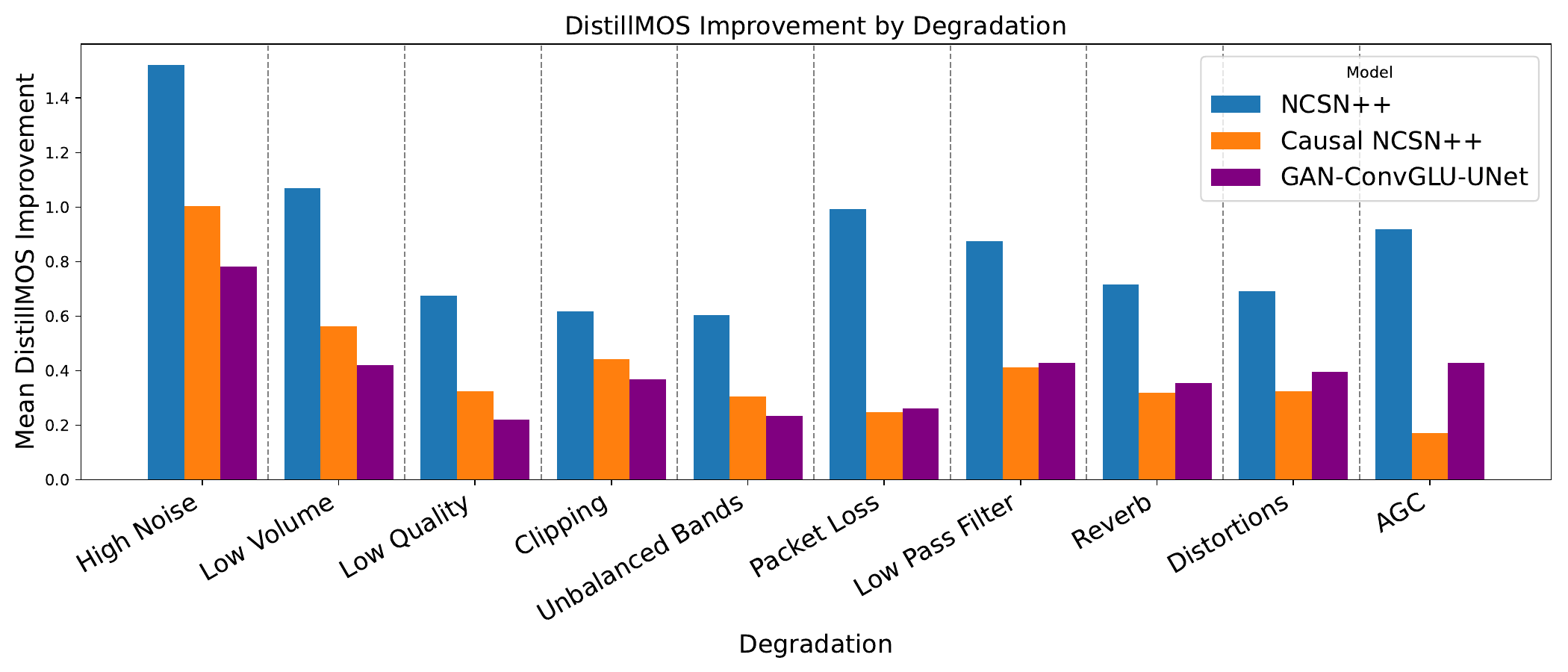}
    \caption{Per-degradation DistillMOS improvement on SIG testset.}
    \vspace{-10pt}
    \label{fig:per_deg}
\end{figure}
\subsection{Per-degradation Performance Comparison}
We visualize per-degradation DistillMOS improvements for the FM- and GAN-based models in Fig.~\ref{fig:per_deg}. The degradation types have been annotated internally on the SIG testset by at least 3 raters per file. Degradation categories are sorted in descending order by the difference in mean improvement between the causal NCSN++ and the GAN-ConvGLU-UNet.
Relative to GAN-ConvGLU-UNet, the causal NCSN++ is stronger on high noise, low volume, low quality, clipping, and unbalanced bands, and is comparable on packet loss, whereas GAN surpasses the causal NCSN++ on low-pass filtering, reverberation, generic distortions, and AGC. 
% This pattern aligns with our broader finding \textcolor{blue}{findings}

\section{Conclusion}
\label{sec:conclusion}
We explored real-time causal flow-matching for speech restoration operating in the compressed complex STFT domain. We evaluate two different model architectures both requiring only 20 ms total latency. 
We find that introducing the latency constraints significantly affects performance compared to the established non-causal NCSN++ baseline. We show that it is possible to use simpler and less complex architectures for FM, but performance seems tightly connected to the compute effort. We observe that weaker models benefit from longer sampling trajectories, but too long trajectories introduces degradation and hallucination. Using the same architecture, FM performs equal or worse than a comparable GAN baseline. This study is a first step to indicate where FM is under realistic real-time constraints, but shows that more work is needed to achieve satisfactory performance and matching existing techniques. 

\bibliographystyle{IEEEbib-abbrev} % for abbreviating author names
\bibliography{refs}

\begin{thebibliography}{10}

\bibitem{loizou2007speech}
P.~C. Loizou,
\newblock {\em Speech enhancement: theory and practice},
\newblock CRC press, 2007.

\bibitem{o2024speech}
D. O'Shaughnessy,
\newblock ``Speech enhancement—a review of modern methods,''
\newblock {\em IEEE Transactions on Human-Machine Systems}, vol. 54, no. 1, pp. 110--120, 2024.

\bibitem{narayanan2013ideal}
A. Narayanan and D. Wang,
\newblock ``Ideal ratio mask estimation using deep neural networks for robust speech recognition,''
\newblock in {\em IEEE International Conference on Acoustics, Speech and Signal Processing}, 2013, pp. 7092--7096.

\bibitem{erdogan2015phase}
H. Erdogan, J.~R. Hershey, S. Watanabe, et~al.,
\newblock ``Phase-sensitive and recognition-boosted speech separation using deep recurrent neural networks,''
\newblock in {\em IEEE International Conference on Acoustics, Speech and Signal Processing}, 2015, pp. 708--712.

\bibitem{williamson2015complex}
D.~S. Williamson, Y. Wang, and D. Wang,
\newblock ``Complex ratio masking for monaural speech separation,''
\newblock {\em IEEE/ACM Transactions on Audio, Speech, and Language Processing}, vol. 24, no. 3, pp. 483--492, 2015.

\bibitem{tan2018convolutional}
K. Tan and D. Wang,
\newblock ``A convolutional recurrent neural network for real-time speech enhancement,''
\newblock in {\em Proc. Interspeech}, 2018, pp. 3229--3233.

\bibitem{luo2019conv}
Y. Luo and N. Mesgarani,
\newblock ``Conv-tasnet: Surpassing ideal time--frequency magnitude masking for speech separation,''
\newblock {\em IEEE/ACM Transactions on Audio, Speech, and Language Processing}, vol. 27, no. 8, pp. 1256--1266, 2019.

\bibitem{hu2024general}
Q. Hu, T. Tan, M. Tang, et~al.,
\newblock ``General speech restoration using two-stage generative adversarial networks,''
\newblock in {\em IEEE International Conference on Acoustics, Speech, and Signal Processing Workshops}, 2024, pp. 31--32.

\bibitem{lemercier2023analysing}
J.-M. Lemercier, J. Richter, S. Welker, et~al.,
\newblock ``Analysing diffusion-based generative approaches versus discriminative approaches for speech restoration,''
\newblock in {\em IEEE International Conference on Acoustics, Speech and Signal Processing}, 2023, pp. 1--5.

\bibitem{richter2023speech}
J. Richter, S. Welker, J.-M. Lemercier, et~al.,
\newblock ``Speech enhancement and dereverberation with diffusion-based generative models,''
\newblock {\em IEEE/ACM Transactions on Audio, Speech, and Language Processing}, vol. 31, pp. 2351--2364, 2023.

\bibitem{ho2020denoising}
J. Ho, A. Jain, and P. Abbeel,
\newblock ``Denoising diffusion probabilistic models,''
\newblock {\em Advances in Neural Information Processing Systems}, pp. 6840--6851, 2020.

\bibitem{song2020score}
Y. Song, J. Sohl-Dickstein, D.~P. Kingma, et~al.,
\newblock ``Score-based generative modeling through stochastic differential equations,''
\newblock {\em Proc. International Conference on Learning Representations}, 2021.

\bibitem{lu2022conditional}
Y.-J. Lu, Z.-Q. Wang, S. Watanabe, et~al.,
\newblock ``Conditional diffusion probabilistic model for speech enhancement,''
\newblock in {\em IEEE International Conference on Acoustics, Speech and Signal Processing}, 2022, pp. 7402--7406.

\bibitem{welker22speech}
S. Welker, J. Richter, and T. Gerkmann,
\newblock ``Speech enhancement with score-based generative models in the complex {STFT} domain,''
\newblock in {\em Proc. Interspeech}, 2022, pp. 2928--2932.

\bibitem{lay2024single}
B. Lay, J.-M. Lermercier, J. Richter, et~al.,
\newblock ``Single and few-step diffusion for generative speech enhancement,''
\newblock in {\em IEEE International Conference on Acoustics, Speech and Signal Processing}. IEEE, 2024, pp. 626--630.

\bibitem{ristea2024icassp}
N.~C. Ristea, A. Saabas, R. Cutler, et~al.,
\newblock ``Icassp 2024 speech signal improvement challenge,''
\newblock in {\em IEEE International Conference on Acoustics, Speech and Signal Processing}, 2024.

\bibitem{lee2025flowse}
S. Lee, S. Cheong, S. Han, et~al.,
\newblock ``Flowse: Flow matching-based speech enhancement,''
\newblock in {\em IEEE International Conference on Acoustics, Speech and Signal Processing}, 2025, pp. 1--5.

\bibitem{jung2024flowavse}
C. Jung, S. Lee, J.-H. Kim, et~al.,
\newblock ``Flowavse: Efficient audio-visual speech enhancement with conditional flow matching,''
\newblock {\em arXiv preprint arXiv:2406.09286}, 2024.

\bibitem{liu2024generative}
A.~H. Liu, M. Le, A. Vyas, et~al.,
\newblock ``Generative pre-training for speech with flow matching,''
\newblock in {\em Proc. International Conference on Learning Representations}, 2024.

\bibitem{ku2025generative}
P.-J. Ku, A.~H. Liu, K. Korostik, et~al.,
\newblock ``Generative speech foundation model pretraining for high-quality speech extraction and restoration,''
\newblock in {\em IEEE International Conference on Acoustics, Speech and Signal Processing}. IEEE, 2025, pp. 1--5.

\bibitem{korostik2025modifying}
R. Korostik, R. Nasretdinov, and A. Juki{\'c},
\newblock ``Modifying flow matching for generative speech enhancement,''
\newblock in {\em IEEE International Conference on Acoustics, Speech and Signal Processing}. IEEE, 2025, pp. 1--5.

\bibitem{richter2024causal}
J. Richter, S. Welker, J.-M. Lemercier, et~al.,
\newblock ``Causal diffusion models for generalized speech enhancement,''
\newblock {\em IEEE Open Journal of Signal Processing}, vol. 5, pp. 780--789, 2024.

\bibitem{lipman2022flow}
Y. Lipman, R.~T. Chen, H. Ben-Hamu, et~al.,
\newblock ``Flow matching for generative modeling,''
\newblock in {\em International Conference on Learning Representations}, 2022.

\bibitem{tong2024improving}
A. Tong, K. Fatras, N. Malkin, et~al.,
\newblock ``Improving and generalizing flow-based generative models with minibatch optimal transport,''
\newblock {\em Transactions on Machine Learning Research}, 2024.

\bibitem{braun2022effect}
S. Braun and H. Gamper,
\newblock ``Effect of noise suppression losses on speech distortion and asr performance,''
\newblock in {\em IEEE International Conference on Acoustics, Speech and Signal Processing}, 2022.

\bibitem{dubey2022icassp}
H. Dubey, V. Gopal, R. Cutler, et~al.,
\newblock ``Icassp 2022 deep noise suppression challenge,''
\newblock in {\em IEEE International Conference on Acoustics, Speech and Signal Processing}, 2022.

\bibitem{radford2023robust}
A. Radford, J.~W. Kim, T. Xu, et~al.,
\newblock ``Robust speech recognition via large-scale weak supervision,''
\newblock in {\em International Conference on Machine Learning}, 2023, pp. 28492--28518.

\bibitem{morris2004and}
A.~C. Morris, V. Maier, and F.~D. Green,
\newblock ``From wer and ril to mer and wil: improved evaluation measures for connected speech recognition.,''
\newblock in {\em Proc. Interspeech}, 2004, pp. 2765--2768.

\bibitem{reddy2021dnsmos}
C.~K. Reddy, V. Gopal, and R. Cutler,
\newblock ``Dnsmos: A non-intrusive perceptual objective speech quality metric to evaluate noise suppressors,''
\newblock in {\em IEEE International Conference on Acoustics, Speech and Signal Processing}, 2021.

\bibitem{stahl2025distillation}
B. Stahl and H. Gamper,
\newblock ``Distillation and pruning for scalable self-supervised representation-based speech quality assessment,''
\newblock in {\em Proc. IEEE International Conference on Acoustics, Speech and Signal Processing ({ICASSP})}, 2025.

\bibitem{reddy2021interspeech}
C.~K. Reddy, H. Dubey, K. Koishida, et~al.,
\newblock ``Interspeech 2021 deep noise suppression challenge,''
\newblock {\em Proc. Interspeech}, 2021.

\bibitem{defossez2022highfi}
A. Défossez, J. Copet, G. Synnaeve, et~al.,
\newblock ``High fidelity neural audio compression,''
\newblock {\em Transactions on Machine Learning Research}, 2022.

\end{thebibliography}

\end{document}